\documentstyle[10pt,aaspp4,epsf,rotate]{article}

\def\psim{\lower.5ex\hbox{$\; \buildrel \propto \over \sim \;$}}
\def\u{{\upsilon }}
\def\d{{\delta}}
\def\e{{\epsilon}}
\begin{document}

\title{Fireball Loading and the Blast Wave Model of Gamma Ray Bursts}

\author{Charles D. Dermer\altaffilmark{1}, James Chiang\altaffilmark{1,2}, and Markus
B\"ottcher\altaffilmark{1,3}}

\altaffiltext{1}{E. O. Hulburt Center for Space Research, Code 7653,
       Naval Research Laboratory, Washington, DC 20375-5352}
\altaffiltext{2}{NRL/NRC Resident Research Asscociate}
\altaffiltext{3}{Department of Space Physics and Astronomy, Rice University, Houston, TX 
77005-1892}

\begin{abstract} 

A simple function for the spectral power $P(\epsilon,t) \equiv \nu L(\nu) $
is proposed to model, with 9 parameters, the spectral and temporal evolution
of the observed nonthermal synchrotron power flux from GRBs in the blast
wave model. Here $\epsilon = h\nu/$m$_e$c$^2$ is the observed dimensionless photon
energy and $t$ is the observing time.  Assumptions and an issue of lack
of self-consistency are spelled out.  The spectra are found to be most sensitive to the 
baryon loading, expressed in terms of the initial bulk Lorentz factor $\Gamma_0$, and
an equipartition term $q$ which is assumed to be constant in time and independent of
$\Gamma_0$.  Expressions are given for the peak spectral power $P_p(t) = 
P(\epsilon_p,t)$ at the photon energy  $\epsilon  =
\epsilon_p(t)$ of the spectral power peak. A general rule is that the total
fireball particle kinetic energy $E_0 \sim \Pi_0 t_d$, where $t_d \propto \Gamma_0^{-8/3}$ is
the deceleration time scale and $\Pi_0 \equiv P(\epsilon_p,t_d) \propto \Gamma_0^{8/3}$ is
the maximum measured bolometric power output in radiation, during which it is carried
primarily by photons with energy ${\cal E}_0 = \epsilon_p(t_d) \propto q\Gamma_0^4$. 

This rule governs the general behavior of fireballs with different baryon loading.  Clean
fireballs with small baryon loading ($\Gamma_0\gg 300$) are intense, subsecond,
medium-to-high  energy
$\gamma$-ray events, and are difficult to detect because
of deadtime and sensitivity limitations of previous $\gamma$-ray
detectors such as EGRET on {\it CGRO}. Dirty fireballs
with large baryon loading ($\Gamma_0\ll 300$) produce transient
emissions which are longer lasting and most luminous at X-ray energies and
below,  but these events are lost behind the glow of the X-ray and
lower-energy background radiations except for rare serendipitous detections
by pointed instruments.  The correlation between hardness and duration of 
loaded GRB fireballs ($100\lesssim\Gamma_0\lesssim 1000$) follows from this rule.

\end{abstract}

\section*{1. Introduction}

The cosmological origin of GRBs has been established as a result of follow-up
observations of fading X-ray counterparts to GRB sources discovered with the {\it
Beppo-SAX} mission (e.g., Costa et al.\ \markcite{cea97}1997; Metzger et 
al.\ \markcite{mea97}1997; van Paradijs et al.\ \markcite{vea97}1997; 
Djorgovski et al.\ 
\markcite{dea97}1997; Kulkarni et al.\ \markcite{kea98a}1998a).  The time
profiles of the X-ray afterglow light curves are generally well fit by
power laws with temporal indices $\chi$ in the range $1.1\lesssim \chi
\lesssim 1.5$ where, at a fixed observing frequency, the flux $\phi\propto t^{-\chi}$ 
(e.g, Costa et al.\ \markcite{cea97}1997;  Feroci et
al.\ \markcite{fea98}1998). Only the X-ray afterglow behavior of GRB 970508
is not well fit with a single power law, as its X-ray flux displays a peak
between 1-2 days after the GRB. Its overall X-ray behavior from 30 seconds to
1 week following the burst event does, however, follow a power-law
decay with $\chi = 1.1$ (Piro et al.\ \markcite{pea98a}1998a). The light curves of
GRB counterparts detected at optical and radio frequencies also generally display power-law behavior
with a similar range of temporal indices (e.g., Sahu et al.\ \markcite{sea97}1997; Galama
et al.\ \markcite{gea98}1998; but note Groot et al.\ \markcite{grea98}1998; Frail
\markcite{Frail98}1998).

The long wavelength afterglows of GRBs are most simply explained with the
fireball/blast-wave model, and were predicted prior to
their discovery (Paczy\'nski \& Rhoads \markcite{pr93}1993; M\'esz\'aros \& Rees
\markcite{mr97}1997).  In this model, the impulsive release of a large amount of energy in a small
volume is transformed into relativistic plasma whose baryon-loading parameter,
expressed in terms of the initial bulk Lorentz factor $\Gamma_0$ of the
fireball after it has become optically thin, obtains values much greater
than unity (Blandford
\& McKee \markcite{bm76}1976; Cavallo \& Rees \markcite{cr78}1978; Shemi \&
Piran \markcite{sp90}1990; Rees \& M\'esz\'aros \markcite{mr92}1992; Piran \&
Shemi \markcite{ps93}1993). The dominant radiation mechanism producing the
prompt GRB emission and radio through X-ray afterglows is nonthermal synchrotron
emission (M\'esz\'aros \& Rees \markcite{mr93a}1993a; Katz
\markcite{katz94}1994; Tavani \markcite{tavani96}1996; Cohen et al.\
\markcite{cea97}1997). The power-law behavior of the long wavelength afterglows is a
consequence of the deceleration of the blast wave as it sweeps up material from the
ambient medium (Wijers, Rees, \& M\'esz\'aros \markcite{wrm97}1997; Waxman
\markcite{waxman97a}1997a; Vietri
\markcite{vietri97a}1997a; Dermer \& Chiang \markcite{dc98}1998).  The
blast-wave model also provides an explanation for the $> 100$ MeV emission
observed with the EGRET instrument from bright BATSE bursts (Dingus
\markcite{dingus95}1995; Catelli, Dingus, \& Schneid
\markcite{catelli96}1996; Hurley et al.\ \markcite{hurley94}1994) as due
either to synchrotron self-Compton (SSC) emission (M\'esz\'aros,  Rees, \& Papathanassiou
\markcite{mrp94}1994; Chiang \& Dermer \markcite{cd98}1998) or to the
nonthermal proton synchrotron radiation from hadrons accelerated in the relativistic
blast wave (Vietri \markcite{vietri97b}1997b; B\"ottcher \& Dermer 1998).

In this paper, we propose a simple function to describe the observed prompt
and delayed emissions from a decelerating blast wave in the
fireball/blast-wave model. We use this function to perform a parameter
study, and find that the temporal and spectral behavior of the blast-wave
emissions are robust to orders-of-magnitude changes in all parameter values
except for the baryon-loading parameter $\Gamma_0$ and an equipartition parameter $q$. 
If $q$ is not strongly dependent on $\Gamma_0$, as quantified below, then the
observability of a fireball is most strongly determined by the value of the
baryon-loading factor
$\Gamma_0$.  Clean fireballs with small baryon loading ($\Gamma_0\gg 300$)
produce intense, subsecond, medium-to-high energy ($\gg 10$ MeV)
$\gamma$-ray events, and heavily-loaded dirty fireballs  ($\Gamma_0\ll 300$)
produce longer duration transients which are most luminous at X-ray energies
and below.  Limitations of earlier telescopes may have prevented
the discovery of clean fireballs, though some previous serendipitous detections
of X-ray transients might represent members of the dirty fireball class. 

In Section 2, we introduce a function to model the temporally-evolving nonthermal
synchrotron emission produced by relativistic blast waves which decelerate and radiate
by sweeping up particles from the circumburst medium (CBM). The terms in this function
are derived from basic blast-wave physics. Differences in the emission properties of
blast waves due to changes in parameter values are examined in Section 3. In
particular, the strong-dependence of the blast-wave spectra on $\Gamma_0$ is noted. 
In Section 4, we consider the limitations of and requirements for telescopes to be
capable of detecting clean and dirty fireballs in view of the emission properties of
blast waves with different baryon loading. The predictions of clean and
dirty fireballs as new classes of astrophysical transients are discussed in Section 5,
where the central results of this study are summarized through a simple rule which governs
the overall emission properties of blast waves produced by relativistic fireballs.

\section*{2. Spectral Power and Spectral Power Flux of GRBs in the Blast Wave Model}

GRB emissions are most conveniently described in terms of their measured $\nu F_\nu$  
spectral power fluxes.  For simplicity, we consider only uncollimated, spherically
expanding blast waves here.  The spectral power $P(\epsilon,t) = \nu L_\nu$ is related
to the spectral power flux through the relation $ P(\epsilon,t) =  4\pi d_L^2 \nu
F_\nu$, where $d_L$ is the luminosity distance.  Due to the strong Doppler beaming when
$\Gamma
\gg 1$, most of the observed flux is produced by emission regions located within an
angle
$\theta
\lesssim 1/\Gamma$ of the line-of-sight direction.  As the blast wave decelerates into
and through the nonrelativistic  regime, the observer receives additional contributions
from emission regions at larger angles to the line-of-sight (Panaitescu
\&  M\'esz\'aros \markcite{pm98a}1998a; Waxman
\markcite{waxman97b}1997b; Sari \markcite{sari98}1998; Chiang \& Dermer
\markcite{cd98}1998).  This introduces corrections to the above relation between the produced
power and observed flux, but this effect is not important when $\Gamma \gg 1$, and is
neglected here. 

In this limit, we model the temporally-evolving GRB spectrum with the function
$$P(\e,t) \;\left[\epsilon\times {{\rm ergs}\over{\rm s}-\epsilon }\right]\;=\; {(1+\u/\delta)\;
P_p(t)\over [\e/\e_p(t)]^{-\u}+(\u/\delta)[\e/\e_p(t)]^{\d}  }\;\eqno(1)$$ 
where, on a $\nu L_\nu$ plot, the rising slope has index $\u$ ({\it up}silon), and the 
descending slope has index $\delta$ ($\delta${\it own}). Both
$\u$ and $\delta$ are $> 0$. The dimensionless photon energy $\epsilon$ is in units of $m_e$, where 
particle masses are in energy units.  Obviously 
$P(\epsilon_p,t) = P_p(t)$. Expression (1) applies only to the nonthermal synchrotron
portion of the spectrum, in accord with detailed treatments (Chiang \& Dermer
\markcite{cd98}1998; Panaitescu
\& M\'esz\'aros \markcite{pm98b}1998b) which show that the
SSC contribution to the total power output is generally much weaker than the nonthermal
electron synchrotron contribution.  The SSC component can still represent the dominant
emission process in certain frequency ranges, for example, at $\gg 100$ MeV
gamma-ray energies or at soft gamma-ray energies in the afterglow phase when the
high-energy flux is at a very low level. At MeV energies and below, however, the SSC component can
usually be neglected.  Compton processes involving external photons can also be shown to
be negligible unless the surrounding radiation fields have extremely large energy
densities $\gtrsim 10^4$ eV cm$^{-3}$ (see, e.g., Dermer,  Sturner, \& Schlickeiser
\markcite{dss97}1997).

We assume that the density of the CBM surrounding the location of the
burst event can be parameterized by a power law in distance from the center of the 
explosion.  Thus
$$n(x)  = n_0 x^{-\eta}\;.\eqno(2)$$
The dimensionless spatial coordinate $x$ represents the distance from the explosion
site in units of the deceleration radius
$$x_d = [{(3-\eta)E_0\over 4\pi n_0\Gamma_0^2 m_p}]^{1/3}\;,\eqno(3)$$ 
which characterizes the distance at which the blast wave has 
swept up $\approx 1/\Gamma_0$ times the initial baryonic mass of the fireball  (Rees \&
M\'esz\'aros \markcite{rm92}1992; M\'esz\'aros \& Rees \markcite{mr93b}1993b).

It is useful to parameterize the evolution of the bulk
Lorentz factor of the blast wave by the expression 
$$\Gamma(x) = \cases{\Gamma_0~,& if $0 \leq x < 1$;\cr\cr
  \Gamma_0 x^{-g}~,& if $1\leq x \leq \Gamma_0^{1/g}.$\cr}\eqno(4)$$  The index
$g$ refers to the deceleration regime.  If very little of the swept-up energy  is
radiated over the time scale on which the blast wave decelerates, then the blast
wave evolves in the non-radiative (or adiabatic regime) with $g\rightarrow 3/2$ when
$\eta = 0$ (Blandford \& McKee
\markcite{bm76}1976; Paczy\'nski \& Rhoads
\markcite{pr93}1993). This has a self-similar solution independent of $\Gamma_0$ when
$x \gg 1$. If, on the other  hand, the bulk of the swept-up energy is promptly
radiated away, then the blast wave evolves in the radiative regime with $g
\rightarrow 3$ when $\eta = 0$ (e.g., M\'esz\'aros,
Rees, \& Wijers \markcite{mrw97}1997 and references therein). Regimes
intermediate to these two limits have been treated analytically (Dermer \& Chiang
\markcite{dc98}1998) and numerically (Chiang \& Dermer \markcite{cd98}1998; Panaitescu
\& M\'esz\'aros \markcite{pm98c}1998c). 

Corresponding to $x_d$ is the deceleration time scale in the observer's
frame, given by
$$t_d = {(1+z)x_d\over \Gamma_0^2 c}\;= {(1+z)\over
c\Gamma_0^{8/3}}\;\left[{(3-\eta) E_0\over 4\pi n_0
m_p}\right]^{1/3}\;.\eqno(5)$$  
Eqs.\ (3) and (5) hold when $\eta < 3$, and are
readily generalized for more complicated radial structure involving, for example,
inner and outer boundaries of the CBM, or more general radial
distributions with $\eta = \eta(r)$.  In the most general case, $\eta=\eta({\bf r}$). 
A more complicated radial dependence of $\eta$ than given through eq.\ (2) would
produce additional time-dependent variations in the spectral output of the
decelerating blast wave than derived here. For example, short time scale variability
in GRB light curves may result from clouds surrounding the GRB source (Dermer \&
Mitman \markcite{dm98}1998). The dimensionless spatial coordinate $x$ is
related to the observed time $t$ through the expression
$$ x =  \cases{t/t_d~,& if $0\leq t < t_d$ and $0 \leq x < 1$;\cr\cr
                 [(2g+1){t\over t_d} -2g]^{1/(2g+1)}~,& if $t_d \leq t \leq 
                { t_d\over ( 2 g + 1) }\, \left( \Gamma_0^{2  + g^{-1}}
                 + 2 \, g \right)$ and $1\leq x \leq \Gamma_0^{1/g}.$\cr}
\eqno(6)$$ 

The existence of a low-energy cutoff in the electron distribution function is
reasonable,  given that the blast wave sweeps up electrons and protons from the
CBM with Lorentz factor $\Gamma$ in the comoving frame.  If an
efficient mechanism transfers energy from the protons to the electrons, then the
low-energy cutoff of the electron Lorentz factors  can reach values as large as
$$\gamma_{\rm e,min} = \xi_e (m_p/m_e)\Gamma \;\eqno(7)$$ (e.g., M\'esz\'aros
et al.\ \markcite{mrp94}1994). The term $\xi_e <
1$ is an electron equipartition factor. Synchrotron radiation from an
electron distribution with a low-energy cutoff produces a low-energy spectrum with $\u =
4/3$ (Katz \markcite{katz94}1994; Tavani \markcite{tavani96}1996; Cohen et al.\
\markcite{cea97}1997).

Fermi processes in the blast wave shock can additionally accelerate a nonthermal
component  of electrons and protons. If the injection index is steeper than 3, or if
the injection index  is steeper than 2 and cooling processes are rapid, then the $\nu
F_\nu$  spectrum falls with $\delta  > 0$.  It is assumed here that the values of $\u$
and $\delta$ and, by implication, the spectrum of the underlying  particle
distributions do not evolve with time. This is clearly not the case due to particle
cooling through radiative processes (see Chiang
\& Dermer \markcite{cd98}1998 for a numerical treament).  These processes must be
carefully considered for interpretations of and fitting to afterglow spectra.  But
these details are not so important for the overall energetics of the blast-wave
emissions considered here.

Another crucial uncertainty of blast-wave models is the magnetic field strength $H$ in the 
comoving blast wave frame.  This is commonly given in terms of an ``equipartition field"
obtained by  equating the magnetic field energy density with the nonthermal particle energy density
of the swept-up  particles downstream of the forward shock.  Thus $H$ is parameterized by the
expression
$$ H ({\rm G}) = [32\pi m_p n(x) \xi_H (r/4)]^{1/2}\Gamma(x) = 116 \; [n(x) \xi_H
(r/4)]^{1/2}  \Gamma_{300}\; ,\eqno(8)$$ 
where $n(x)$ is the density of the swept-up gas in the stationary frame
of the GRB source given by eq.\ (2), $r$ is the compression ratio,
$\Gamma_{300}= \Gamma(x)/300$, and $\xi_H$ is the magnetic-field equipartition
parameter.   Field generation through dynamo processes could strengthen $H$ with time
(see, e.g., Ryu \& Vishniac \markcite{rv91}1991; M\'esz\'aros, Laguna, \& Rees
\markcite{mlr93}1993 and references therein);  flux-freezing and reconnection could
weaken it (e.g., M\'esz\'aros et al.\ \markcite{mrp94}1994).  The choice $\xi_H \sim 
1$ does not produce GRB spectral forms similar to those observed, as excessive cooling 
produces a spectral component with $\u \cong 0.5$ (Sari, Piran, \& Narayan
\markcite{spn98}1998; Chiang \& Dermer
\markcite{cd98}1998).  Values of $\xi_H\lesssim  10^{-4}$ are required to give good fits
to GRB spectra during the prompt phase.  Although there is no justification for
treating $\xi_H$ as time-independent when $\xi_H\neq 1$, we do so here for simplicity.

In a tangled field with mean field strength $H$ given by eq.\ (8), the low-energy
electron  Lorentz-factor cutoff (7)  gives a peak in the spectral power at energy
$H\gamma^2_{\rm e,min}/H_{\rm cr}$ in the comoving frame, where $H_{\rm cr} =
4.414\times 10^{13}~{\rm G}$.   To the observer, this energy is boosted and
redshifted by a factor $\approx
\Gamma/(1+z)$.  Thus  the observed peak energy
$\epsilon_p$ at time $t$ is given by
$$\epsilon_p(t) = {\cal E}_0[\Gamma(x)/\Gamma_0]^4 x^{-\eta/2} = {\cal E}_0
\cases{x^{-\eta/2} \;, & if $0\leq x < 1$;\cr\cr
                      x^{-4g-\eta/2}\;,& if $1\leq x < \Gamma_0^{1/g}$,\cr}\eqno(9)$$  
where $${\cal E}_0 ={ 3.0\times 10^{-8} n_0^{1/2} q \Gamma_0^4\over (1+z)} \; ,
\eqno(10)$$
 and $$q \equiv [\xi_H (r/4)]^{1/2}\xi_e^2\;.\eqno(11)$$
The coefficient in eq.\ (10) is
$(m_p/m_e)^2 (32\pi m_p)^{1/2}/H_{\rm cr} = 2.97\times 10^{-8}$, and ${\cal E}_0$
represents the observed photon energy of the peak of the $\nu F_\nu$ spectrum when the
blast wave passes through the deceleration radius at $x = 1$.

The power in swept-up particle kinetic energy in the comoving frame is
$$\dot E_{\rm ke} = m_p \, c \, B(x) \, \Gamma(x) \, [\Gamma(x) - 1] \, n(x) \, A(x)
\eqno(12)$$ (Blandford \& McKee \markcite{bm76}1976), where $A(x) = A_0 x^2 =
(4\pi x_d^2) x^2$ is the area  of the spherically expanding blast wave and $B(x) c = [1 -
\Gamma(x)^{-2}]^{1/2} \, c$ is its speed.  If $\zeta$ is defined as the fraction of
swept-up energy retained in the comoving frame, then $(1-\zeta)$ is the
fraction that is dissipated by the blast wave shock. Although this dissipation could be
in the form of energetic particles which escape directly
from the shock, or in the form of bulk kinetic energy which is randomized and stored in the
reverse shock to be radiated away later or used to accelerate the blast wave front (see, e.g., 
M\'esz\'aros \& Rees \markcite{1997c}1997; Panaitescu \& M\'esz\'aros
\markcite{pm98a}1998a,\markcite{pm98c}c), we assume that essentially all of the dissipation
occurs in the form of photon radiation. 

As previously  discussed, only the emitting region along the line of sight makes a
significant  contribution to the observed flux in the limit
$\Gamma\gg1$.  The observed power is boosted and redshifted by the factor
$\Gamma^2/(1+z)^2$ over the radiant power in the comoving frame.  Consequently we obtain an
expression relating the observed power to the swept-up power, given by 
$$\dot E_{\rm rad} \cong {\Gamma^2\over (1+z)^2}(1-\zeta ) \dot E_{\rm ke} \cong  (1-\zeta)m_p
c\Gamma^4 n(x) A(x)/ (1+z)^2\; .\eqno(13)$$ 
The spectral power given by eq.\ (1) satisfies the normalization 
$$E_0 = \int_0^\infty dt\;\dot E_{\rm rad}(t) =  \int_0^\infty dt \int_0^\infty \;d\epsilon
\;\epsilon^{-1} P(\epsilon,t)\; ,\eqno(14)$$  where $E_0$ is the total energy in baryons at the
end of fireball coasting phase. Thus the radiated power
$$\dot E_{\rm rad} \cong 2(\upsilon^{-1} +\delta^{-1}) P_p(t) \; .\eqno(15)$$

We use the approach of Dermer \&
Chiang (\markcite{dc98}1998) to find that in the regime
$1\ll x \ll \zeta \Gamma_0^{1/g}$, the fraction of swept-up energy that is dissipated
promptly is given by $(1-\zeta) = (2g-3+\eta)/g $ (see Appendix), and  we assume that this
relation holds elsewhere as well. Consequently we derive an expression for the measured bolometric
spectral power, given by 
$$ P_p(t) =  \Pi_0\cases{x^{2-\eta},& if $0\leq x < 1 \; $ ;\cr\cr
                      x^{2-\eta -4g},& if $1\leq x < \Gamma_0^{1/g}\; $ .\cr}\eqno(16)$$ 
The coefficient 
$$\Pi_0 = {(2g-3+\eta) m_p c\Gamma_0^4 n_0 A_0\over 2g(\u^{-1}+\delta^{-1})(1+z)^2}\;= 
{c(2g-3+\eta)\Gamma_0^{8/3} \over 2g(\u^{-1}+\delta^{-1})(1+z)^2}\;(4\pi m_p
n_0)^{1/3}(3-\eta)^{2/3} E_0^{2/3}\; \eqno(17)$$ represents the bolometric luminosity
(ergs s$^{-1}$) at the time $t_d$ of peak power output. We again note that a more
complicated expression would hold if $\eta$ were not constant, thereby rendering $\zeta$
dependent upon time.

It is straightforward to determine the time-dependence of $\epsilon_p(t)$ and
$P_p(t)$ in different regimes from eqs.\ (6), (9), and (16).  For the parameterized
density distribution (2), the peak photon energy $\epsilon_p(t) \propto t^{-\eta/2}$ 
when
$t\leq t_d$, whereas for 
$t \ge t_d$, $\epsilon_p(t) \propto t^{-(12+\eta)/8}$ and
$\propto  t^{-(24+\eta)/14}$ in the non-radiative and radiative regimes, respectively.
For a uniform medium ($\eta = 0$),  $\epsilon_p (t)$ is constant
for $t\leq t_d$, whereas  for $t \geq t_d$, $\epsilon_p(t)
\propto t^{-3/2}$ and $\propto t^{-12/7}$ in the non-radiative and radiative regimes,
respectively. The peak spectral power $P_p(t)$ increases $\propto t^{2-\eta}$ for 
$t\leq t_d$, and $P_p(t) \propto t^{-1-\eta/4}$ and $\propto t^{-(10+\eta)/7}$  in the
limiting non-radiative ($g=3/2$) and radiative ($g=3$) regimes, respectively, when $t
\gg t_d$. For a uniform medium with $\eta = 0$,  $P_p (t)\propto t^2$
for $t\leq t_d$, and $P_p(t)
\propto t^{-1}$ in the non-radiative regime when $t \geq t_d$ (Sari
\markcite{sari97}1997). In the radiative regime, $P_p(t)\propto t^{-10/7}$ when $t \geq
t_d$. From these relations, one can see that the power emitted in lower-energy radiation such
as the optical band will always be less than in the higher energy X-ray and gamma-ray
bands for declining density profiles (i.e., $\eta > 0$).

One can also easily determine the temporal indices measured at a fixed observing  energy
$\epsilon$ by consulting equations (1), (6), (9), and (16). Let  $t_p(\epsilon)$
represent the time when the spectral power flux at
$\epsilon$ reaches its maximum value.  It is found by solving
$\epsilon = \epsilon_p(t_p)$ using eqs.\ (6) and (9), yielding
$$ t_p (\epsilon) = {t_d \over 2 g + 1} \, \left[ \left( {\epsilon \over {\cal E}_0} 
\right)^{-{2 g + 1 \over 4 g + \eta/2}} + 2 \, g \right]  . \eqno(18) $$
Note that  $t_p(\e ) \rightarrow t_d$ when $\epsilon \gg{\cal E}_0$, whereas
$t_p(\e )\gg t_p$ when $\epsilon \ll{\cal E}_0$.  The displacement of the pulse peaks
at lower energies may act alone or in concert with synchrotron cooling (Chiang
\markcite{chiang98}1998; Dermer \markcite{dermer98}1998) to produce the observed energy dependence
of the GRB pulse width (Fenimore et al.\ \markcite{fea95}1995; Piro et al.\
\markcite{pea98b}1998b).  Effect (18) represents a prediction of this model, and will be explored
in more detail elsewhere (Dermer, B\"ottcher, \& Chiang
\markcite{dbc98}1998).

Two regimes characterize the time-dependence of the flux at high energies $\epsilon
\gg{\cal E}_0$, namely
$$ P(\epsilon,t) = (1+{\delta\over \u})\Pi_0 ({\epsilon\over {\cal E}_0 })^{-\delta}
\;\times  \cases{ (t/t_d)^{2 - \eta - \eta\delta/2}, & if $t \ll t_d$, \cr\cr 
[(1+2g)t/t_d]^{[2 - \eta (1 +
\d / 2) - 4g ( 1 +  \d )] /(2g+1)}, & if $t \gg t_d$. \cr}
\eqno(19)$$
In contrast, three regimes characterize the time-dependence of the flux at energies
$\epsilon \ll{\cal E}_0$. They are
$$  P(\epsilon,t) = \Pi_0 
\;\times   \cases { (1+\u /\d) ({\epsilon\over {\cal E}_0 })^{\u}(t/t_d)^{2 - \eta +
\eta\u /2}, & if $t \ll t_d$,
\cr\cr  (1+\u /\d) ({\epsilon\over {\cal E}_0 })^{\u}[(1+2g)t/t_d]^{[2 - \eta (1-\u
/2) - 4g(1-\u )]/(2g+1)}, & if $t_d \ll t \ll t_p(\e ) $, \cr\cr
(1+ \d / \u) ({\epsilon\over {\cal E}_0 })^{-\d }[(1+2g)t/t_d]^{[2 - \eta (1 +
\d / 2) - 4g ( 1 +  \d )] /(2g+1)}, & if $t \gg t_p(\e )$. \cr}
\eqno(20)$$
Note that the late time behaviors in eqs.\ (19) and (20) are identical.

At times $t > t_d$, two types of behavior occur depending on whether the observations
take place at $\e \ll {\cal E}_0 $ or $\e \gg {\cal E}_0 $. In the former
case, there are two branches in the time profile related to the time $t_p(\epsilon)$
when $\e_p$ sweeps through the observing energy $\e $. When $t_d \ll t \ll t_p(\e )$, 
$$ P(\epsilon,t) \propto t^{\chi_1} \hskip 0.5cm {\rm where} \hskip 0.5cm 
\chi_1 = [2 - \eta (1-\u /2) - 4g(1-\u )]/(2g+1)\;.
\eqno(21)$$ 
When $t\gg t_p(\e )$, the time profile decays according to the relation
$$ P(\epsilon,t) \propto t^{-\chi_2} \hskip 0.5cm {\rm where} \hskip 0.5cm
\chi_2 = [4g ( 1 +  \d ) + \eta (1 + \d / 2) - 2] /(2g+1)
\eqno(22)$$
In the latter high-energy case ($\e \gg {\cal E}_0 $), a single power-law decay in the
time profile is found for $t \gg t_d$ which follows the behavior given by eq.\ (22).
Table 1 lists the derived temporal slopes $\chi_1$ and $\chi_2$ for various values of
$\u,\delta$, and $g$. 

\section*{3. Model Spectra}

Table 2 gives a list of the parameters which, when placed in eqs.\ (5), (6), (9),
(10), (16), and (17), can be used to calculate temporally-evolving model spectra using
spectral form (1). Fig.\ 1a shows such a calculation employing the parameter values
listed in Table 2 and  called  ``standard" henceforth. For these parameters, the
deceleration time scale $t_d = 58$ s and 
${\cal E}_0 = 0.4$, giving a duration and a 200 keV $\nu F_\nu$ peak
energy early in the event which are characteristic of typical GRBs. The
smoothly joined broken power-law form resembles the canonical GRB ``Band"-type
spectrum (Band et al. \markcite{Band93}1993). This model burst radiates a peak power
$\Pi_0 \cong 3.8\times 10^{50}$ ergs s$^{-1}$ at $t_d$ and would be below the BATSE
 flux threshold at redshift $z = 5$, but would be strongly detected if the GRB source were
at $z = 1$.

The spectra in Fig.\ 1a are plotted at observing times from 1 $\mu$s to $10^7$ s in
factor-of-10 increments. The solid lines which overplot the spectra connect
the peak energies $\epsilon_p(t)$ and peak luminosities $P_p(t)$ at different
observing times.  When $t\leq t_d$, $\epsilon_p(t)$ is constant because $\eta = 0$
in this example, whereas it declines at later times according to the relation
$\epsilon_p(t)\propto t^{-4g/(2g+1)}\propto t^{-1.6}$, noting eqs.\ (6) and (9) with
$g = 2$.  For these same parameters, the peak power output
$P_p(t)\propto t^2$ when 
$t\leq t_d$, but $P_p(t)$ decreases $\propto t^{(2-4g)/(2g+1)} \propto t^{-6/5}$ when $t \gg t_d$,
noting eqs.\ (6) and (16).  We see that observations of the temporal decay of the
$\nu F_\nu$ peak relate the indices $g$ and $\eta$ which characterize the radiative
regime and density profile of the external medium, respectively. Again, the possible
spatial dependence of $\eta$ complicates any simple interpretation of the measured
behavior.

Fig.\ 1b is a representation of the model data shown in Fig.\ 1a in the form of time profiles
at different observing energies. Light curves are shown at $\epsilon = 10^{-12},\; 10^{-9},\;
10^{-6},\; 10^{-3},\; 1,\;  10^{3},$ and $10^6 $, roughly corresponding to radio, mm, optical,
X-ray, MeV, GeV, and TeV frequencies, respectively. The light curves at $t > t_d$ display one or two
power-law behaviors depending on whether  $\epsilon > {\cal E}_0 = 0.4$ or $\epsilon <
0.4$, respectively, for the reasons given in the previous section. For the
very late time behavior with $t
\gg t_p(\e ) > t_d$, the temporal decay index $\chi_2 = [4g(1+\delta)-2]/(2g+1) = 1.52$, in agreement
with the behavior shown in Fig.\ 1b.  It is interesting to note that temporal indices
in the observed range $1.1\lesssim \chi \lesssim 1.5$ require flat spectra with
$\delta \lesssim 1/3$ when
$\eta = 0$ (see Table 1) .

The GRB spectral indices $\u$ and $\d$ are assumed for simplicity not to evolve with
time.  Thus spectra at photon energies $\e \ll \e_p$ and $e\gg
\e_p$ rise and fall with the indices $\u = 4/3$ and $\d = 0.2$, respectively, for the
parameters given in Table 2.  To examine the effects of different parameter values on
the observed spectra, it is therefore sufficient to plot the $P_p(t)$-$\e_p(t)$
trajectory at various times, as indicated in Fig.\ 1a by the straight lines.  Figs.\
2a-e presents a parameter study in this form.  The Table 2 standard is shown in each
of these figures by the solid lines with open circles at times incremented by
factors-of-10 starting from 1 $\mu$s.

Fig.\ 2a shows the effects of changing the index which characterizes the radiative regime.  Here
we show two cases with $g = 1.6$ and $g = 2.9$, in addition to the $g = 2$ standard.
The case $g = 1.6$ is near the non-radiative limit (if
$g=3/2$, there is no radiation by definition), and $g = 2.9$ is near the fully radiative limit.
The weakly radiative blast wave with $g = 1.6$ persists for a longer period of time, though it
radiates less power in its early phases. The highly radiative blast wave with $g =
2.9$ is nearly an order-of-magnitude more luminous in its early phases than the
$g=1.6$ case, but its inertia drops more quickly and it therefore has a shorter
luminous life than in the less radiative cases.  

At this point, a remark concerning a serious issue regarding lack of
self-consistency of the parameter list is in order.  The type of radiative regime
specified by $g$ essentially gives the fraction of swept-up power that is dissipated.
If nonthermal synchrotron radiation represents the dominant power drain from the blast
wave, as we assume, then a fraction $\approx 2-g^{-1}(3-\eta)$ of the incoming swept-up power is
promptly transformed into radiation.  This places  demands on the underlying particle
distribution and comoving energy densities which are not considered here, but must be kept in mind
in any analytic treatment employing a parameterization of blast-wave dynamics according to an
expression like eq.\ (4). A detailed numerical simulation with a self-consistent treatment of
plasmoid/blast-wave dynamics is essential to model spectra accurately (Chiang \& Dermer
\markcite{cd98}1998).

Figs.\ 2b and 2c show the effects of varying the density $n_0$ and the explosion
energy
$E_0$, respectively, over four orders of magnitude. In higher density
environments, a blast wave's energy is radiated more rapidly and, provided $q$ is
the same, in the form of higher energy photons; thus its peak power output obtains
larger values than in more dilute environments.  Fireballs with larger total energies
but with equal baryon-loading persist for longer periods of time because both  $t_d$
and the time scale to decelerate to nonrelativistic energies are $\propto E_0^{1/3}$
(see eqs.\ [5] and [6]). For a fixed baryon loading factor (or constant $\Gamma_0$),
the overall duration of the GRB is multiplied by a factor
$\propto  (E_0/n_0)^{1/3}$ as a consequence of the deceleration time scale (see eqs.[3]
and [5]). This dependence comes from dimensional analysis in scaling solutions for
spherical hydrodynamics, and is present in both the Sedov solution (see, e.g.,
Lozinskaya \markcite{loz92}1992) and its relativistic equivalent derived by Blandford
\& McKee  \markcite{bm76}1976).   The peak power output changes to compensate for the
different duration; thus  $P_p(t_d)\propto E_0^{2/3}$ when $E_0$ varies and
$n_0$ is held constant, and
$P_p(t_d)\propto n_0^{1/3}$  when $n_0$ varies and $E_0$ is constant (eq.\ [17]).
This latter dependence is a function of the radiative efficiency, which is assumed to be
identical to the factor $(1-\zeta)$ derived in the Appendix.  Consequently, the
dimensional dependence of $P_p(t)$ is more complicated than that of $t_d$. In any case, the
durations and powers are not strongly dependent on changes of $n_0$ and $E_0$.

Fig.\ 2d shows the effect of varying the equipartition term $q$ over four orders-of-magnitude,
from $q = 10^{-5}$ to $q = 10^{-1}$.  A change in $q$ by a constant factor causes the curves to be
displaced horizontally by the same constant factor, since $\e_p(t)\propto q$ but $P_p(t)$ is
independent of $q$. 

The effects of different amounts of baryon loading are shown in Fig.\ 2e. The spectra
are found to be extremely sensitive to changes in  $\Gamma_0$.  The mean photon
energy ${\cal E}_0=\e_p(t_d)$ at the time of peak-power output
$\Pi_0$ is $\propto \Gamma_0^4$.  Thus most of the radiation of a clean fireball with
$\Gamma_0 = 3000$ is in the form of photons with energies eight orders of magnitude
larger than the emission from a dirty fireball with
$\Gamma_0 = 30$, provided that $q$ remains constant. The duration of peak power output
is just the deceleration timescale $t_d$ which is  $\propto \Gamma_0^{-8/3}$. To
compensate for the constant total energy in the fireball, $\Pi_0 \propto
\Gamma_0^{8/3}$. Thus the duration of peak power output of the clean fireball case is a factor
$\sim 2\times 10^5$ shorter than that of the dirty fireball case in this example, but
the magnitude of the power emitted during this period is a factor of $\sim 2\times
10^5$ greater.

Clean fireballs produce extremely short pulses of radiation which are carried by very high-energy
photons. For example, when  $\Gamma_0 = 3000$, the peak power output of
$\approx 10^{53}$ ergs is carried  by $\gtrsim 1$ GeV photons in a burst lasting $\lesssim 0.1$ s. 
Again, the total energy  remains roughly constant, though it is carried by many
fewer, though much higher-energy photons. Its late-time behavior approaches that of
fireballs with smaller values of
$\Gamma_0$, but with lower power to compensate for the energy radiated during
the early phases of the event.  For the dirty fireball with $\Gamma_0 = 30$, the peak
power output at UV energies lasts for
$\sim 10$ minutes and develops into a broadband radiation pulse which evolves to lower
energies at later times.  Incidentially, this behavior is  found in blazar flares (e.g., Marscher \&
Gear \markcite{mg85}1985), though a closer examination must  be made to see if this model
provides an adequate explanation for such phenomena. 

The GRBs which trigger BATSE display a very broad bimodal duration distribution ranging from $\sim 30$
ms to $\approx 500$ seconds (Kouveliotou et al.\ \markcite{kea93}1993; Belli
\markcite{Belli95}1995).  A short, hard component of GRBs is seen with durations ranging from
$\approx 30$ ms to 1 second, with a mean duration of   $\approx 0.2$ seconds. A longer duration
component ranges from $\approx 2$-$500$ seconds, with a mean duration of 15 - 30 seconds. 
This hardness-duration correlation is in accord with the behavior
outlined above for blast waves with different baryon-loading, where the class of
short, hard bursts are produced by fireballs which are less baryon-loaded than
average. A model for GRBs in terms of a distribution of $\Gamma_0$ is needed,
however, to establish whether these GRBs constitute a separate class, or an extension
of the typical loaded fireball in the cleaner limit (Dermer et al.
\markcite{dbc98}1998). Different levels of baryon-loading might be expected if GRBs
originate from multiple sources, as indicated if the claimed association of GRB 980425
with SN 1998bw is correct (Kulkarni et al.\ 1998b). For the purposes of this paper, we
consider a continuous range of fireball-loading parameters rather than different
levels of baryon loading.

Thus we find that the observed spectra from fireball/blast waves, which decelerate
and are energized by the process of sweeping up material from the CBM, are most
strongly dependent on the baryon-loading parameter
$\Gamma_0$ and the equipartition parameter $q$.  The term
$q$ determines the mean photon energy of the radiant emission through the relation ${\cal E}_0 \propto
q\Gamma_0^4$ (eqs.\ [9]-[11]).  In the most general case,
$q = q(\Gamma_0)$, and the observational properties of the fireball would depend on the exact form of
this dependence.  We assume that $q$ is only weakly dependent on $\Gamma_0$, that is, we assume
that $q$ depends on $\Gamma_0$ though a function which varies  much less strongly than $\propto
\Gamma_0^{-4}$. In what follows, we examine the observational signatures of fireballs with
different baryon loading $\Gamma_0$ but with constant $q$, keeping in mind the possibility of an
underlying dependence of $q$ on $\Gamma_0$.

\section*{4. Prospects for Detecting Clean and Dirty Fireballs}

The total number of photons collected from a GRB with energies near the mean photon energy of peak
power output is roughly given by 
$$N_\gamma \approx {\Pi_0 t_d A_{\rm det}\over 4\pi d_L^2 m_e {\cal E}_0}\;,\eqno(23)$$
where
$A_{\rm det}$ is the effective area of the detector at energy ${\cal E}_0$.  The
deceleration time scale (5) defines the rough duration of peak power output,
and eq.\ (10) gives the mean photon energy of the peak power output at $t=t_d$.  Hence
$$t_d{\rm (s)} = 1730\;\left({q\over{\cal E}_0}\right)^{2/3}\; [(1-\eta/3) (1+z)
E_{54} ]^{1/3}\;  \cong \; 31 \; ({q_{-3}\over {\cal
E}_0})^{2/3}\;[(1-\eta/3) \zeta_6 E_{54}]^{1/3}\;,\eqno(24)$$   
where $q_{-3}=q/10^{-3}$ and $\zeta_6 \equiv (z+1)/6$. Recalling the expression for the
magnitude of peak-power output given by eq.\ (17),  we obtain the result
$$N_\gamma \approx { (2g-3+\eta)(3-\eta) E_0 \over 2g(\u^{-1}+\delta^{-1})
(1+z) }\;{ A_{\rm det} \over 4\pi m_e d_L^2 {\cal E}_0}\;\equiv K_\gamma{ E_{54} A_{\rm det}
\over  d_{29}^2 {\cal E}_0(1+z)} ,\eqno(25)$$
\noindent where $d_{29} = d_L/(10^{29} {\rm cm})$. Note that $d_{29} \cong 1$ when $z = 5$
in a $q_0 = 1/2$ cosmology with a Hubble constant equal to $ 65$ km s$^{-1}$ Mpc$^{-1}$.  For
our standard parameters
$\eta = 0$,
$\u = 4/3$, and $\delta = 0.2$,
$$K_\gamma =  \cases {0.32,  & for $g = 1.6$;\cr\cr
                      2.4,& for  $g=2.9$.\cr}\eqno(26)$$

\subsection*{4a. Clean ($\Gamma_0\gg 300$) Fireballs }

The 100 MeV threshold of EGRET corresponds to ${\cal E}_0
\simeq 200$.  To produce 100 MeV nonthermal synchrotron photons mostly from electrons
with energies $\simeq m_e \gamma_{e,{\rm min}}$ requires fireballs with baryon-loading 
$$\Gamma_0 \cong 76 [ { {\cal E}_0 (1+z)\over n_0^{1/2} q }]^{1/4} = 1420
({\zeta_6 \over n_2^{1/2} q_{-3} } )^{1/4} \;,
\eqno(27)$$
where we set ${\cal E}_0 = 200$ on the right-hand-side of eq.\ (27).  The number of photons
detected from a clean fireball, given the standard parameters used in eqs.\ (25) and (26), is
$N_\gamma \approx E_{54}A_{\rm det}[{\rm cm}^2]/[ 3750 d_{29}^2\zeta_6]$ and 
$N_\gamma \approx E_{54}A_{\rm det}[{\rm cm}^2]/[ 490 d_{29}^2\zeta_6 ]$ for $g=1.6$ and
$g=2.9$, respectively. To detect clean fireballs requires that $N_\gamma > 1$. Thus sources
of clean fireballs must be located at distances $d_{29}$ given through the relation 
$$\zeta_6^{1/2}d_{29} \lesssim  E_{54}^{1/2} A^{1/2}_{1000} \cases {0.5  & for $g =
1.6$\cr\cr
                      1.4,& for  $g=2.9$\cr}\;,\eqno(28)$$ where 1000$A_{1000}$ cm$^2$ is the
effective collecting area of the detector at ${\cal E}_0\approx 200$. 

The effective area of EGRET at 100 MeV in the wide field mode (e.g., Kurfess et
al. \markcite{kurfess97}1997) is $\sim 800$ cm$^2$ on-axis and  $\sim 400$ cm$^2$ at 20$^\circ$
off-axis, so clean fireballs would have to be located at $z \ll 5$
and at $z \lesssim 5$ to be detected with EGRET in the non-radiative and radiative limits,
respectively.  Bursts from clean fireballs with ${\cal E}_0
\simeq 200$ would have durations $t_d \approx 0.9 [q^2_{-3}\zeta_6 (1-\eta/3)
E_{54}]^{1/3}$ (eq.\ [24]) at $z \cong 5$ and shorter durations when
$z\ll 5$ and $q_{-3} \ll 1$. The spark chamber of EGRET requires
$\approx 0.1$ s to recover from a photon event, although it can  detect multiple photons if they
arrive within a few microseconds.  Consequently, EGRET is insensitive to more than one
photon arriving between several microseconds and $\sim 100$ ms, which is a critical range for the
clean fireballs considered here.  

It might be surprising that no sub-second bursts of gamma rays have been reported
(Fichtel et al. \markcite{fea94}1994; Fichtel \& Sreekumar \markcite{fs97}1997) from some members
of the clean fireball class which are located at $z\lesssim 5$. We propose three possibilities
to explain why this class of clean fireballs with peak power output near 100 MeV has not been
discovered:
\begin{itemize}
\item There are many fewer clean fireballs, as defined by eq.\ (27), than those which
produce the loaded BATSE GRBs.
\item The duration of clean fireballs is typically $\lesssim 0.1$ s, so that deadtime limitations
have prevented their discovery.
\item Members of the clean fireball class evolve  through cosmic time differently than the loaded
fireballs which produce the GRBs detected with BATSE, and they do so in such a way to have
escaped detection. 
\end{itemize}
 
If the third point does not apply, then we can derive a requirement on the
redshift $z_B$ of the dimmest BATSE bursts to agree with the failure to detect clean fireballs
with EGRET.   Very crudely, we scale the number of clean fireballs which could be detected
with EGRET to the BATSE detection rate ($\sim 800$ yr$^{-1}$ full sky) and the EGRET
field-of-view  ($\sim 1/25$th of the full sky) and lifetime ($\approx 5$ yrs).  If clean
fireballs can only be detected by EGRET within
$z\lesssim z_E$ as given by eq.\ (28), then for EGRET not to have detected at least one member of
this class, we find that
$$k_{cf}\times 0.04 \times \left( {z_E\over z_B}\right)^j\times 800 {\rm ~yr}^{-1}\times 5
{\rm ~yr}  \;\lesssim 1\;.\eqno(29)$$ The term $k_{cf}$ is a class enhancement factor correcting
for the number of clean fireballs relative to the number of loaded fireballs which produce GRBs
detectable with BATSE.  The exponent
$j$ represents a cosmological scaling, and
$j=3$ for Euclidean space. Evolutionary and cosmic expansion effects will introduce
modifications to
$j$ not considered here. Simply taking
$j=3$, we find that $z_E\lesssim 0.2 z_B k_{cf}^{-1/3}$ for the clean fireballs not to have
been detected with EGRET.  Noting eq.\ (28), this implies that even if $k_{cf} \sim 1$,
sensitivity and FoV limitations could have prevented their discovery.  Adding to the
difficulty of the discovery is the deadtime limitations given by the second point above.

Stronger conclusions require a detailed size-distribution study involving the cosmological
evolution of loaded GRB fireballs and clean fireballs in different scenarios, for example, the
hypernova scenario (Paczy\'nski \markcite{Pac98}1998; Wijers et al.\ \markcite{wea98}1998). Given
the uncertainty in the parameters, it remains an open question whether the larger effective area
of the proposed {\it GLAST} instrument will be adequate for the discovery of the clean fireball
class, or if all that is required is a wide FoV, large area gamma-ray calorimeter with fast
timing.  The discovery of
subsecond bursts of $\sim 100$ MeV photons from clean fireballs remains a real possibility
which, should such an event be detected, is not to be confused with the Hawking radiation
from evaporating mini-black holes by virtue of its afterglow behavior.

Clean fireballs would be difficult to detect with the {\it
Solar~Maximum~Mission}, which accumulated on time scales of 16 s and was not sensitive
to short bursts with durations $\lesssim 2$ s  (Harris \& Share \markcite{hs98}1998). The short
hard clean fireball bursts would also be difficult to detect with  COMPTEL on {\it CGRO}, noting
its smaller effective area ($\sim 10^2$ cm$^2$; see, e.g., Hanlon et al.
\markcite{hea95}1995).  Nonetheless, further study of the COMPTEL sample of GRBs detected
with BATSE, and a search for unidentified MeV transients will provide additional constraints
on the number and redshift distribution of  clean ($\Gamma_0\gg 300$) fireballs. 

\subsection*{4b. Dirty ($\Gamma_0\ll 300$) Fireballs }

Because a large number of X-ray telescopes operate in the 1-10 keV range, we can use ${\cal E}_0
\sim 3/511$ to define the dirty fireball regime, which includes those blast waves with
$$\Gamma_0 \simeq 21\;[ { (1+z) \over  n_0^{1/2} q} ]^{1/4}\; \cong 104\; ({\zeta_6 \over
 n_2^{1/2} q_{-3}} )^{1/4}\;
\eqno(30)$$ 
(compare eq.\ [27]). Detection of dirty fireballs with X-ray telescopes
is difficult because of the intense backgrounds.  A signal of
$N_\gamma$ counts given by eq.\ (23) is detected above a background of
$B_G$ counts at the  $n_\sigma$ significance level provided that
$$N_\gamma \gtrsim n_\sigma (2B_G)^{1/2} \; . \eqno(31)$$ The number of background counts
$$B_G \cong \Delta\epsilon_{\rm det}\cdot \Phi_{\rm dif}({\bf \Omega},\epsilon_{\rm
det})\cdot \Delta\Omega\cdot \Delta t \cdot A_{\rm det}\;,\eqno(32)$$ where, from right to
left, the terms are the detector area, the observing time, the field-of-view (FoV) of the
detector, the diffuse photon flux weighted by an average photon energy within the band, and the
bandwidth.  The background noise is composed of instrumental, magnetospheric, zodiacal,
Galactic, and extragalactic components, which are time-dependent in general.  Here we give a
background estimate for X-ray telescopes considering only the diffuse cosmic X-ray
noise background, yielding a lower limit to $B_G$.

In the 3-100 keV range, Boldt (\markcite{Boldt87}1987) gives the expression
$$\epsilon\Phi_{\rm XRB}(\epsilon,\Omega) \;[{{\rm ph} \over {\rm cm}^2 {\rm ~s~sr}}]\;\cong
5.6 \; [{E({\rm keV})\over 3~({\rm keV})}]^{-0.29}\; \exp[-E({\rm keV})/40~{\rm
keV}]\;\eqno(33)$$ for the diffuse extragalactic X-ray background flux. At 3 keV, corresponding
roughly to the mean energy of photons detected by an X-ray telescope sensitive in the 1-10 keV
band, the diffuse flux is $\simeq 5$ ph cm$^{-2}$ s$^{-1}$ sr$^{-1}$.

Substituting eqs.\ (23) and (32) into eq.\ (31) with the stated diffuse flux gives the
result that detection at the $n_\sigma$ level requires an X-ray telescope with 
$${A_{\rm det}\over \Delta \Omega}  \; \left[ {{\rm cm}^2\over {\rm sr}}\right] \gtrsim
{3.4\times 10^{-4}\;d_{29}^4\;n_\sigma^2\; (1+z)^2 \Delta t \over K_\gamma^2 \; E_{\rm 54}^2 \;
f(\Delta t/t_d)}\eqno(34)$$   The function $f(y)$ weights the significance of detection by
the observing interval over the period in which the fireball's emissions are monitored. 
Its form depends on the time-dependence of the light curve of the fireball
emissions, which can be modeled using the eq.\ (1), and the epoch of observation.  In
an illustrative regime, $f(y) = y^2$ for $y \leq 1$, and
$f(y) = 1$ for $y > 1$.  The detection efficiency is clearly optimized when $\Delta t
\approx t_d$. The deceleration time scale for a burst peaking at
${\cal E}_0 \cong 3/511 = 0.006$ is
$$t_d({\rm s}) \simeq 970 \; [\zeta_6 q_{-3}^2 (1-\eta/3) E_{54}]^{1/3}\;.\eqno(35)$$
The seach for dirty fireballs should be conducted in the range of time scales from
$10^2$-$10^4$ s, recognizing the difficulty due to orbital variations of the background
radiation environment.

When $\eta = 0$, one obtains the requirement for an X-ray telescope to be
able to detect a dirty fireball at the $n_\sigma$ level, namely
$${A_{\rm det}\over \Delta \Omega} \gtrsim {\zeta_6^{7/3} d_{29}^4 (n_\sigma /5)^2
q_{-3}^{2/3} \over E_{54}^{5/3} } \cases{940,& for $g = 1.6 $;\cr\cr 125 ,& for
$g=2.9$.\cr}\eqno(36)$$
Typical X-ray detectors have effective areas $\sim 10^2$ cm$^2$ (see Table 3 for
general specifications of some X-ray telescopes used to detect GRBs), so we scale to
a detector of area 100$A_{100}$ cm$^2$.  To detect dirty fireballs above the diffuse
X-ray background means that the detector's FoV must satisfy the relation
$${\Delta\Omega\over 4 \pi} \lesssim {E_{54}^{5/3} A_{100}\over \zeta_6^{7/3} d_{29}^4
q_{-3}^{2/3} (n_\sigma/5)^2   }\;
\cases{0.0027,& for $g = 1.6 $;\cr\cr 0.15 ,& for $g=2.9$.\cr}\eqno(37)$$  Thus an X-ray
detector has to have a FoV $\lesssim 10^\circ\times 10^\circ$ or $\lesssim 80^\circ\times
80^\circ$ in order to detect dirty fireballs at $z\sim 5$, depending on whether the blast wave
decelerates primarily in the non-radiative or radiative regimes, respectively.  Fireballs at
$z\ll 1$ place much weaker constraints on the detector's FoV, as indicated in equation
(37), and then it becomes a question of the number of such sources one could expect to see.

Small FoV instruments will only detect an X-ray flash from a dirty fireball very
rarely unless the class of dirty fireballs is populous. Given that there are $\sim
800$ BATSE-detectable GRBs per year averaged over the full sky, this implies a chance
probability of $\sim 0.02\cdot  k_{df}\cdot k_d^2$ per year for a detector with a
$k_d^\circ\times k_d^\circ$ FoV. The class enhancement factor $k_{df}$ of dirty fireballs
compared to loaded BATSE/GRB fireballs is not known at present, though it can be strongly
constrained by modeling the duration distribution of GRBs detected with BATSE or with
other burst detectors (Dermer et al. \markcite{dbc98}1998). Searches for dirty fireballs
should be made with the Wide Field Camera data on Beppo-SAX, and with future
wide FoV imaging X-ray detectors. Concrete deductions about the probability of detecting
serendipitous dirty fireball X-ray flashes requires a size distribution study tied to a
particular detector's characteristics.  

It is feasible that some of the X-ray flashes
observed in archival searches of {HEAO} (Connors, Serlemitsos, \& Swank
\markcite{cea86}1986) and {\it ROSAT} PSPC (Li et al.
\markcite{lea98}1998;  Sun et al.\ \markcite{sea98}1998) data are emissions from dirty
fireballs only if $k_{df} \gg 1$, given that the effective FoV of the ROSAT PSPC is $\sim
1^\circ\times 1^\circ$. The GRBs detected with {\it Ginga} (Strohmayer et al.
\markcite{stea98}1998) have $\nu F_\nu$ peaks which are on average at lower energies
than those detected with BATSE.  Because {\it Ginga} was sensitive to lower
energies events than BATSE, this might indicate that it is more sensitive to a
population of GRBs with greater baryon-loading.  A straightforward prediction from
this model is that the mean duration of detected GRBs is inversely correlated with the
triggering energy range. 

\section*{5. Discussion}

A slewing strategy such as the one so successfully demonstrated by the Beppo-SAX team and
collaborators (e.g. Costa et al.\ \markcite{cea97}1997; van Paradijs et al.
\markcite{vea97}1997; Piro et al.\ {\markcite{pea98a}1998a,\markcite{pea98b}b) opens rich
possibilities for discoveries which must be carefully weighed in GRB telescope design.
Here we have attempted to provide an exposition of the blast-wave model which provides a
simple way to model the evolving spectral power fluxes, and which furthermore helps us to
understand dedicated and serendipitous GRB observations.  

Eq.\ (1) is the proposed time-dependent spectral form. It employs the 9 parameters listed
in Table 2 that follow from the basic blast-wave model. Its crucial underlying
assumptions are that the photon indices $\u$ and $\delta$ and the equipartition term $q$
are constant in time, and that $q$ is independent or only weakly dependent on $\Gamma_0$.
The energies of photons which carry the bulk of the power are $\propto q\Gamma_0^4$ and are
therefore most strongly dependent on the value of $\Gamma_0$. Over this looms a question
of lack of self-consistency, since specification of the radiative regime
$g$ constrains the magnetic field and photon energy densities and particle distributions to
produce the requisite radiative power. Comparison of eq.\ (1) with the results of a
numerical code employing a self-consistent treatment of plasmoid dynamics (Chiang \&
Dermer \markcite{cd98}1998) indicates that the proposed analytic description of the blast
wave emission spectra remains accurate except when the blast wave is highly radiative.

In spite of these concerns, eq.\ (1) can be easily employed to make estimations of imaging
capabilities of burst detectors given slewing rates of narrow field instruments toward a cosmic
transient. It can be used to calculate photoelectric absorption variations in the GRB
environment and its imprint upon the observed burst spectrum (B\"ottcher et al. 1998). It can
also, of course, be used to model the temporal behavior of GRB spectra and to examine
GRB phenomenology in terms of the fireball/blast-wave model (Dermer et al.\
\markcite{dbc98}1998).

By displaying the spectral power flux in the form of eq.\ (1), we obtained an
important rule on fireball emissions relating the observed peak
power $\Pi_0$,  the characteristic duration $t_d$ of this peak power output, and the mean
energy ${\cal E}_0$ of photons which carry this energy.  The characteristic time scale over
which the bulk of the power is radiated is the deceleration time scale $t_d$ (eq.\ [5]; Rees
\& M\'esz\'aros \markcite{rm92}1992), and
$t_d \propto\Gamma_0^{-8/3}$, where $\Gamma_0$ specifies the baryon loading.  The peak power
output $\Pi_0$ (eq.\ [17]) is  also $\propto \Gamma_0^{8/3}$, assuming that the bulk of
the dissipated energy is in the form of photon radiation.  Because most of the total
particle kinetic energy,
$E_0$, originally in the fireball during its coasting phase is radiated during the
observed time
$t_d$, we have
$$E_0 \approx \Pi_0 t_d\;.\eqno(38)$$ 
Precisely, 
$$\Pi_0 t_d = {(2g-3+\eta) (3-\eta)\over 2g (\u^{-1} +\delta^{-1}) (1+z) }\; E_0 \; = 
\cases{0.033 E_0/(1+z),& for $g = 1.6 $;\cr\cr 0.25  E_0/(1+z) ,& for $g=2.9$.\cr}\eqno(39)$$ 
(In the evaluation on the right-hand-side of this expression, we let $\eta = 0$, $\u =
4/3$, and $\delta = 0.2$.) Moreover, the bulk of this energy is observed in the form of photons
with characteristic energy ${\cal E}_0$ given by eq.\ (10); note that ${\cal E}_0\propto
q\Gamma_0^4$. For these parameters, a full 25\% of the energy of the GRB source is
radiated within the deceleration time scale for a highly radiative fireball.  

Eqs.\ (38) and (39) can be used to summarize the emission properties of fireballs with
different baryon loading and prospects for detecting such fireballs. Fig.\ 3 shows
the variation of $\Pi_0$,  $t_d$, and ${\cal E}_0$ when $\Gamma_0$ ranges from 100 to
1000. The dependence of these observable due to variations in $E_{54}$ and $n_0$ are
also indicated.  The overall powers, time scales, and photon energies of peak power
output of fireballs with $\Gamma_0 \sim 300$, and for clean and dirty fireablls
can be determined from this figure.

We used benchmark characteristics of X-ray, soft
gamma-ray, and medium-energy gamma-ray telescopes to establish criteria for
clean ($\Gamma_0 \gg 300$), loaded ($\Gamma_0 \sim 300$), and dirty ($\Gamma_0 \ll
300$) fireballs.  Detection of clean fireballs at
$\gtrsim$ 100 MeV energies implies $\Gamma_0\gtrsim 1000$ (eq.\ [27]), and the detection of
dirty fireballs at $\sim 3$ keV energies implies
$\Gamma_0\lesssim 100$ (eq.\ [30]).  The clean fireball class was not discovered with EGRET;
this means either that there are relatively few sources of energetic clean fireballs, or
that the powers and durations of clean fireballs are such that sensitivity and
deadtime limitations of previous $> 100$ MeV $\gamma$-ray detectors have prevented
their detection. 

Here we qualify our use of eq.\ (38) which gives the general
observational properties of fireballs with different baryon loading parameters, EGRET has, of course, detected
$> 100$ MeV gamma-ray emission from GRBs detectable with BATSE from the loaded ($100
\lesssim \Gamma_0 \lesssim 1000$) fireball class.  Many of these sorts of GRBs will
undoubtedly be detected with the {\it GLAST} mission as well.  But the $>100$ MeV and
GeV photons detected to date (e.g., Hurley et al.
\markcite{hea94}1994; Dingus \markcite{Dingus95}1995)  are not those produced by the
electrons near the low-energy cutoff of the electron distribution function.  These
photons are much more likely to be the SSC emission (e.g., Chiang \& Dermer
\markcite{cd98}1998) or the nonthermal synchrotron radiation from ultra-high energy
protons (Vietri \markcite{Vietri97b}1997b; B\"ottcher \& Dermer
\markcite{bd98}1998) accelerated in the loaded fireball blast wave.  These emissions
are distinguished by their slow decay, which explains why EGRET often detects this
emission long after the main portion of the BATSE burst has decayed.  What EGRET has
not discovered but is implied by the fireball/blast-wave model is a class of clean
fireballs which produces a luminous subsecond burst of $\gg 100$ MeV radiation. The
prediction of such a discovery with improved $\gamma$-ray telescopes is made here.

We also used analyses outlined by eq.\ (38) to describe the properties
of dirty fireballs. Because $\Pi_0$ and ${\cal E}_0$ are lower in dirtier fireballs, the
diffuse sky backgrounds make detection of such a class of objects by wide-field instruments
problematic.  On the other hand, the expected rate of detection of dirty fireballs by
pointed instruments with a few square degree FoV is very unlikely unless the rate of dirty
fireball explosions greatly exceeds the rate of loaded BATSE/GRB fireballs. Whether any
members of this class has been discovered is uncertain, though X-ray flashes have indeed been
detected in the {\it HEAO} and {\it ROSAT} data bases. We predict that moderate FoV
instruments ($\sim 1$\% of the full sky) with surveying capabilities will discover dirty
fireball X-ray transients with typical durations of
$10^2$-$10^4$ seconds. 

Beppo-SAX has shown that improved imaging and rapid response are key to making progress
in solving the GRB mystery.  Granted this success, the design of a new GRB mission should
therefore employ a hard X-ray/soft GRB detector and a narrow FoV X-ray telescope
which can rapidly slew to the GRB position. But to discover the sister classes of clean and
dirty fireballs implied by the fireball/blast-wave model will require additional X-ray and
gamma-ray capabilities as outlined above and considered in more detail elsewhere (Dermer et
al.\ \markcite{dbc98}1998).

\acknowledgements

The work of CD was supported by the Office of Naval Research and the {\it Compton Gamma Ray
Observatory} Guest Investigator Program.  CD thanks Jim Kurfess, Neil Johnson, Mark
Strickman, and Gerry Share for enlightening discussions, and the referee for a
useful critique.  The work of JC was performed while he held a National Research
Council - Naval Research Laboratory Associateship. MB acknowledges support by  the
German Academic Exchange Service (DAAD). 

\section*{Appendix A. The Fraction of Radiated Energy}

An expression is derived in the asymptotic regime $x \gg x_d$ which relates the
fraction $\zeta$ of swept-up energy retained in the blast wave to the index $g$ which
characterizes the radiative regime and  evolution of the Lorentz factor $\Gamma$ time
scale of the blast wave (see eq.\ [4]).  We start from the equation of momentum
conservation (eq.\ [6] of Dermer \& Chiang [\markcite{dc98}1998]) and note that the
energy in swept-up particles accumulated up to the point $x$ is equal to 

$$ \int\limits_0^{\infty} d p \> \gamma \, N_p (\gamma) = \int\limits_0^x d\tilde x \> n_{\rm
ext} (\tilde x)
\, A (\tilde x) \left[ \zeta \Gamma (\tilde x) + (1 - \zeta) \right]. 
\eqno({\rm A1}) $$ In the asymptotic regime $x_d \ll x \ll x_{\rm tr}$, where 
$x_{\rm tr} = x_d \, \Gamma_0^{1/g}$ and $\Gamma(x) = 
\Gamma_0 (x / x_d)^{-g}\gg 1$, this yields
$$ - {\Gamma' (x) \over \Gamma^2 (x)} = {g \over x \, \Gamma (x)}
\approx {n_0 \, A_0 \, \left( {x \over x_d} \right)^{2 - \eta} 
\over N_{\rm th} + n_0 \, A_0 \, \left[ \Gamma_0 \, \zeta 
\int\limits_{x_d}^x d\tilde x \> \left( {\tilde x \over x_d} 
\right)^{2 - g - \eta}  + (1 - \zeta) \int\limits_0^x d\tilde x 
\> \left( {\tilde x \over x_d} \right)^{2 - \eta} \right] }.
\eqno({\rm A2}) $$ Unless the blast wave is in the extreme radiative regime, the
denominator on the right-hand side of eq.\ (A2) is dominated by the first integral.
Thus if $x \ll
\zeta x_{\rm tr}$, eq.\ (A2)  becomes

$$ {g \over \Gamma_0 \, x_d} \, \left( {x \over x_d} \right)^{g - 1}
\approx {3 - g - \eta \over \zeta \, \Gamma_0 \, x_d} \, 
\left( {x \over x_d} \right)^{g - 1}, 
\eqno({\rm A3}) $$ or

$$ \zeta = {3 - g - \eta \over g}\; . \eqno({\rm A4}) $$

\vskip0.2in

\begin{table}[h]
\small
\caption{Slopes of the Temporal Evolution of the Flux at Fixed Photon Energy when
$t\gg t_d$ and $\eta=0$}
\bigskip
\begin{tabular}{c c c c c c c c c}
\tableline
\tableline slope & g & $\u = 1$ & $\u = 4/3$ \\
\tableline
$\chi_1$ & 1.6 & 0.48 & 0.98 \\
$\chi_1$ & 2.9 & 0.29 & 0.86 \\
\tableline
 &  & $\delta = 0$ & $\delta = 1/5$ & $\delta = 1/3$ 
      & $\delta = 1/2$ & $\delta = 3/4$ & $\delta = 1$ & $\delta = 2$ \\
\tableline
$\chi_2$ & 1.6 & 1.05 & 1.35 & 1.56 & 1.81 & 2.\markcite{}19 & 2.57 & 4.09 \\
$\chi_2$ & 2.9 & 1.41 & 1.75 & 1.98 & 2.26 & 2.69 & 3.12 & 4.82 \\
\tableline
\end{tabular}
\label{table1}
\end{table}

\vskip0.2in

\begin{table}[h]
\small
\caption{Standard Parameters for Model of Evolving GRB Spectral Power Flux}
\bigskip
\begin{tabular}{c l l c}
\tableline\tableline Parameter  & Standard Value & Description \\
  &  \\
\tableline
$\Gamma_0$ & 300 & initial bulk Lorentz factor \\
$q$ & $10^{-3}$ & equipartition term \\
$g$ & 2 & index of $\Gamma$ evolution \\
$E_0$  & $10^{54}$ ergs & total particle kinetic energy in GRB \\
$\u$ & 4/3 & spectral index of rising portion of $\nu L_\nu$ spectrum \\
$\delta$ & 0.2 & spectral index of falling portion of $\nu L_\nu$ spectrum \\
$z$ & 5 & cosmological redshift \\
$n_0$  & $10^2$ cm$^{-3}$ & density at $ x=x_d$ \\
$\eta$  &  0 & index of density distribution\\
\tableline
\end{tabular}
\label{table1}
\end{table}

\eject

\begin{table}[h]
\caption[]{Characteristics of Operating Missions with GRB Detectors}
\begin{flushleft}
\begin{tabular}{lccccc}
\tableline
\tableline
\noalign{\smallskip}
\multicolumn{1}{l}{ } & 
\multicolumn{1}{c}{${\rm Energy~band~(keV)}$} &
\multicolumn{1}{c}{${\rm FoV}$} & 
\multicolumn{1}{c}{${\rm Effective~Area~(cm}^2)$} &
\multicolumn{1}{l}{Sensitivity$~^b$} & 
\multicolumn{1}{c}{Imaging}\\
\noalign{\smallskip}
\hline
\noalign{\smallskip} CGRO BATSE & $50-300$ & $4\pi$ & 1800 cm$^2$$~^a$  & $ 6\times 10^{-8}$$~^c$ 
& few degrees
\\ SAX GRBM& $40-700$ & $20^\circ \times 20^\circ$ & $120$ &  &  $10^\prime\rightarrow
3^\prime$ \\ SAX WFC& $1.5-26$ & $20^\circ \times 20^\circ$ & $120$ & $\sim 10^{-10}$ in $10^3$
s$~^d$ &
$10^\prime\rightarrow 2^\prime$
\\ SAX NFI LECS& $0.1-10$ & $0.5^\circ$ & $22@0.25$ keV & $\sim 3\times 10^{-14}$ in $10^5$ s &
$50^{\prime\prime}$
\\ SAX NFI MECS& $0.1-10$ & $0.5^\circ$ & $150@6$ keV &  &  \\
\noalign{\smallskip}
\hline
\end{tabular}
\end{flushleft}
\end{table}
\noindent$^a$Each Large Area Detector\\
\noindent$^b$ergs cm$^{-2}$ s$^{-1}$\\
\noindent$^v$Large Area Detectors; 10 s GRB\\
\noindent$^d$Beppo-SAX data from homepage; see also Piro et al.\ \markcite{pea98}(1998)

\setcounter{figure}{0}
\begin{figure}
\hbox{
\epsfysize=7.5cm
\epsffile[150 50 600 500]{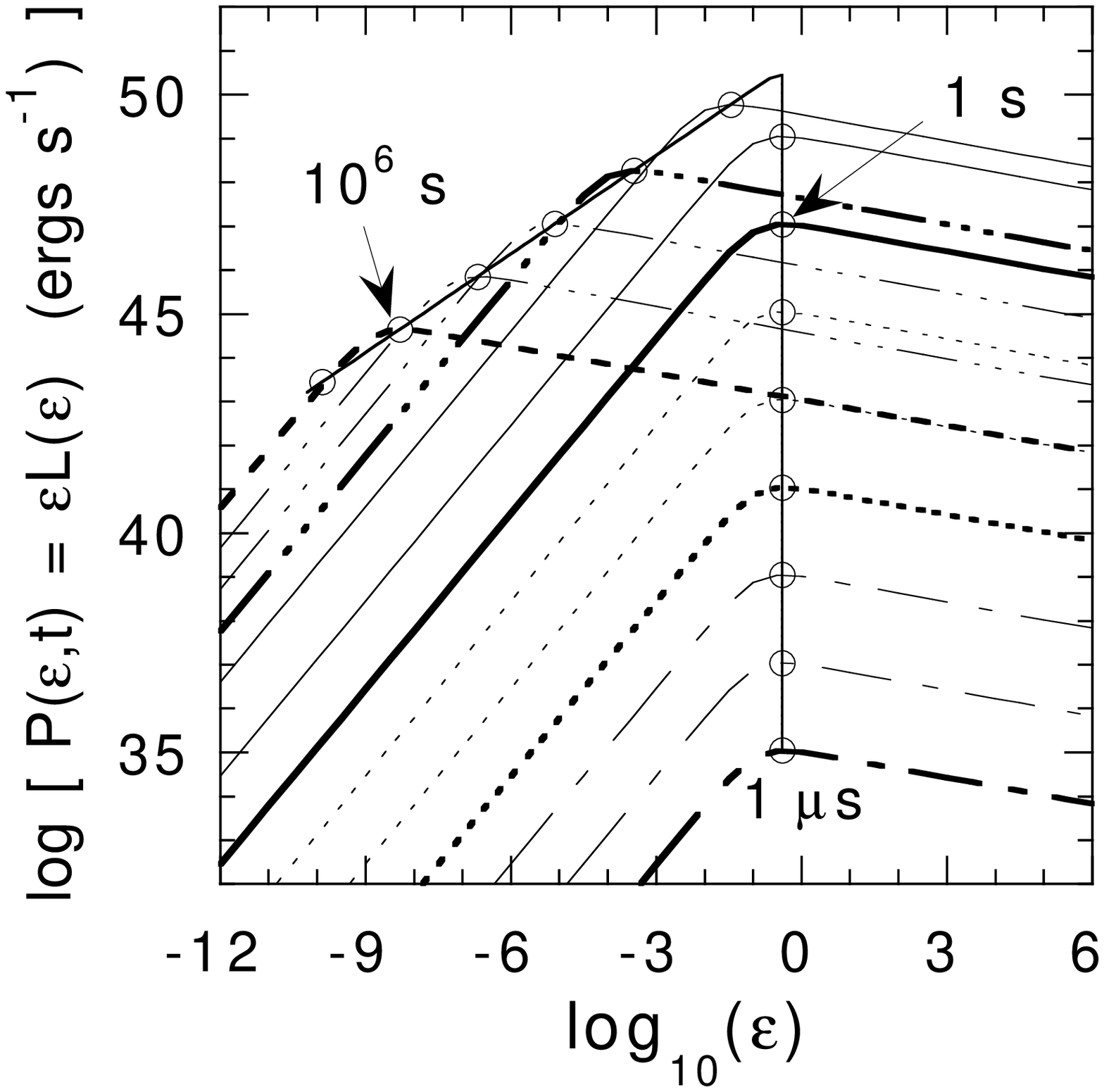} \hskip 1.4cm \epsfysize=7.5cm
\epsffile[150 80 550 500]{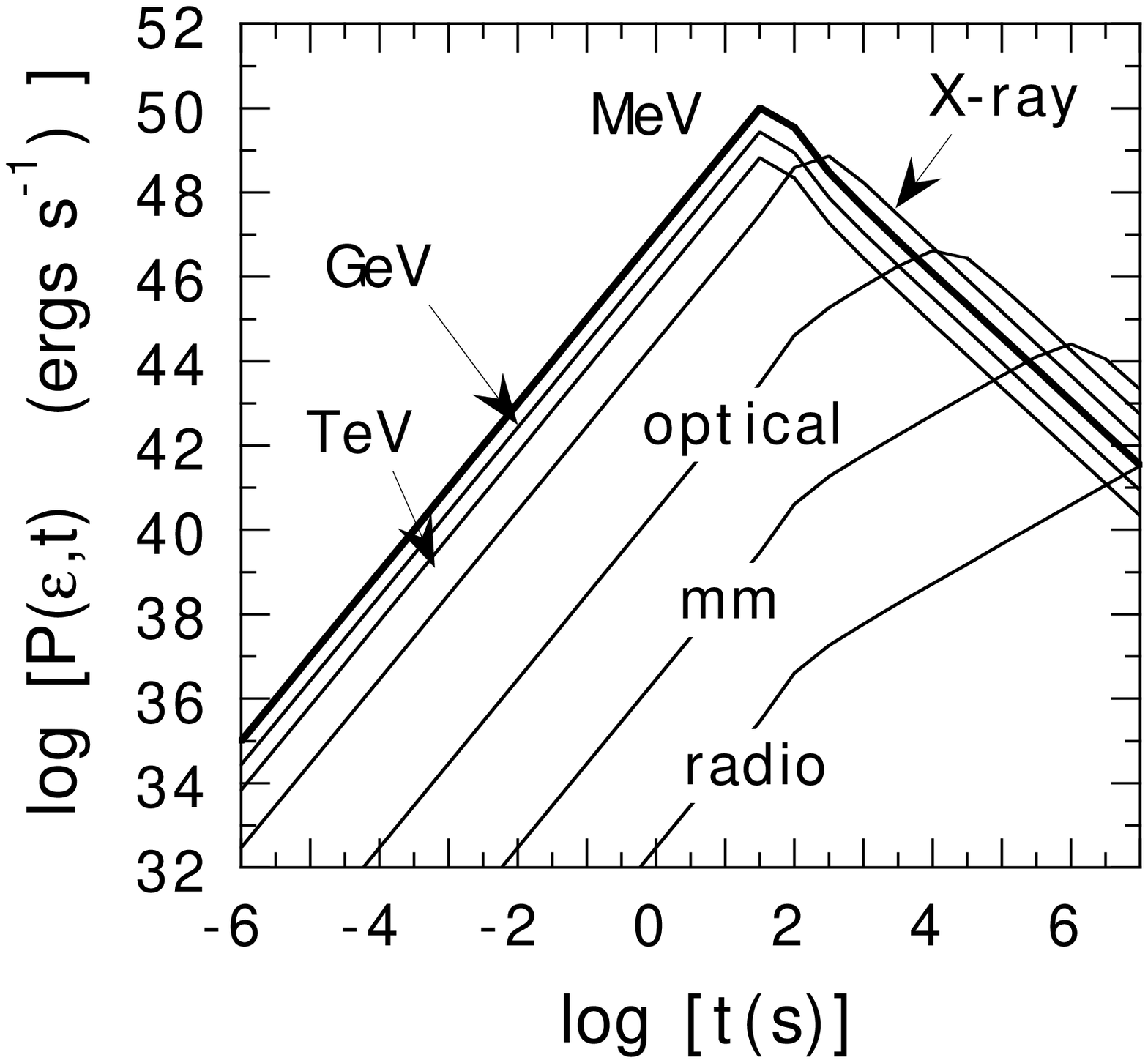}}
\caption[] {
Spectra and light curves for a model GRB using the standard parameters in
Table 2.  (a) Evolution of model GRB spectra at different observing times from 1
$\mu$s to $10^7$ s are shown in factor-of-10 increments. Solid lines plot the
trajectory of the peak power $P_p(t)$ as a function of the photon energy $\e_p(t)$ of
the peak of the
$\nu L_\nu$ spectrum at different observing times. (b) GRB light curves at different
observing energies for the model GRB spectra shown in (a).

}
\end{figure}

\setcounter{figure}{1}
\begin{figure}
\epsfysize=12cm
\epsffile[0 20 500 500]{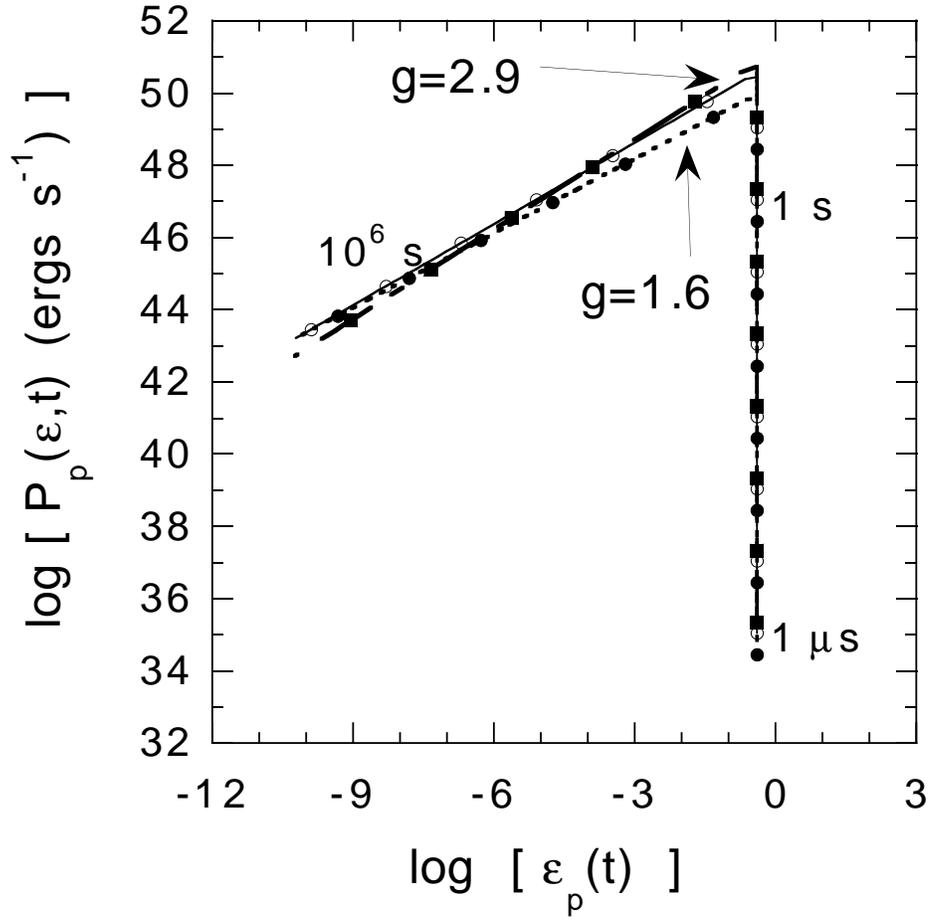}
\caption[]{
Trajectories of model GRBs in the $P_p(t)$-$\e_p(t)$ plane. All
parameters are given in Table 2, except as noted.  (a)
Trajectories for different values of the index $g$ characterizing the radiative
regime and evolution of $\Gamma$ through eq.\ (4). 

}
\end{figure}

\setcounter{figure}{1}
\begin{figure}
\epsfysize=12cm
\epsffile[100 20 500 500]{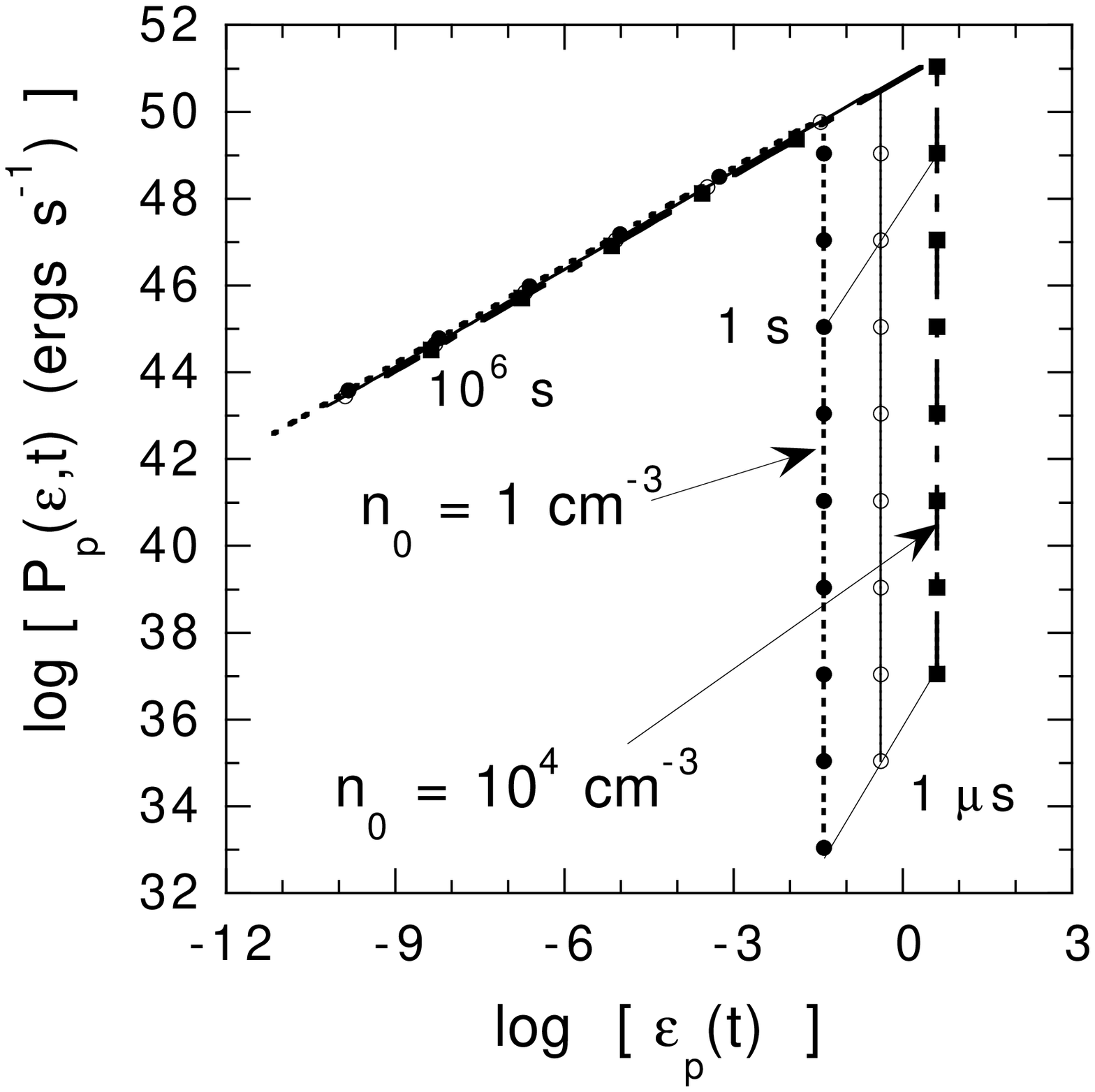}
\caption[]{
(b) Same as Fig.\ 2a, but for
different values of the density $n_0$ of the circumburst medium. }
\end{figure}

\setcounter{figure}{1}
\begin{figure}
\epsfysize=12cm
\epsffile[0 20 500 500]{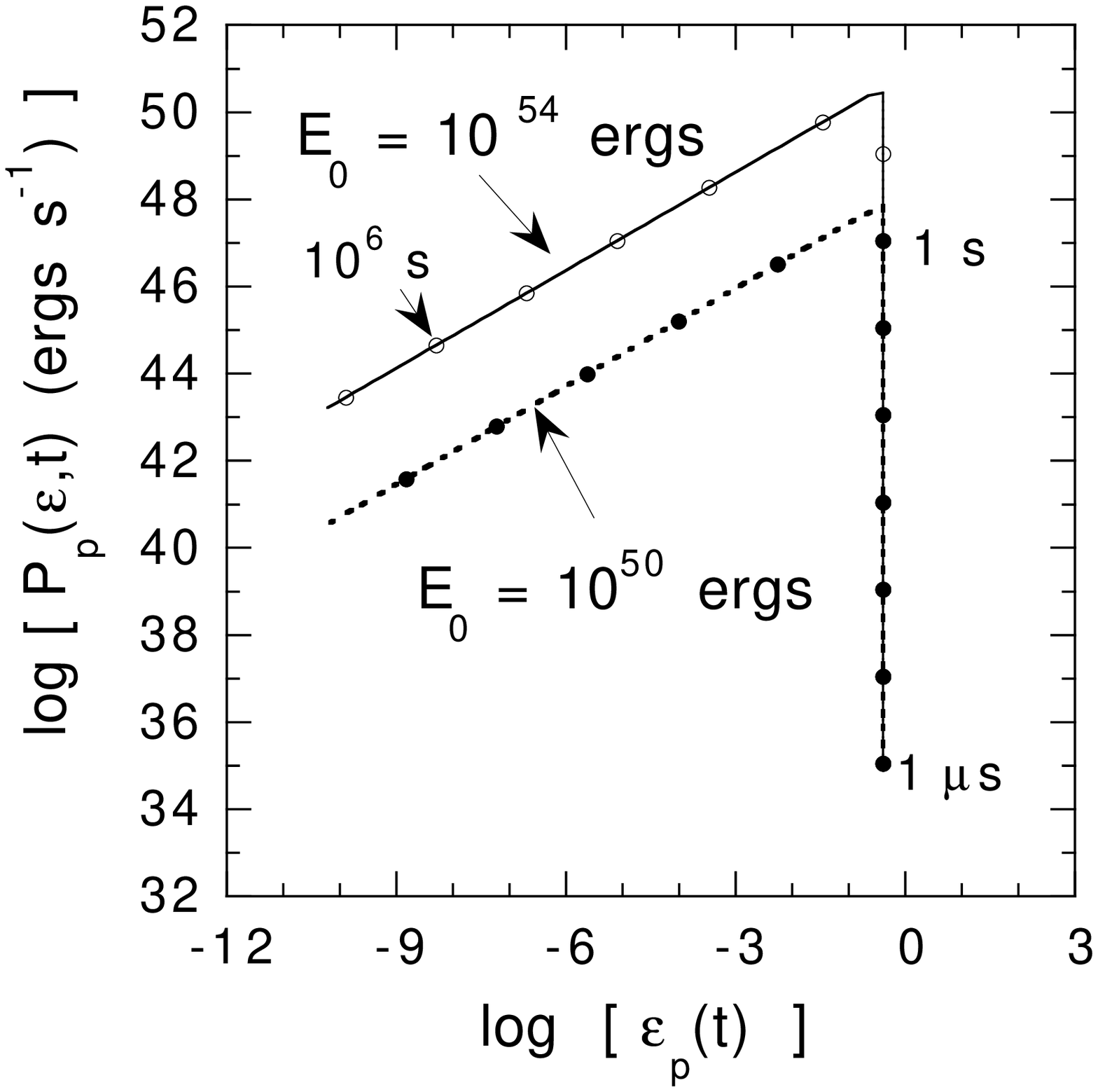}
\caption[]{(c) Same as Fig.\ 2a, but for different values of the total
explosion energy $E_0$.
}
\end{figure}

\setcounter{figure}{1}
\begin{figure}
\epsfysize=12cm
\epsffile[0 20 500 500]{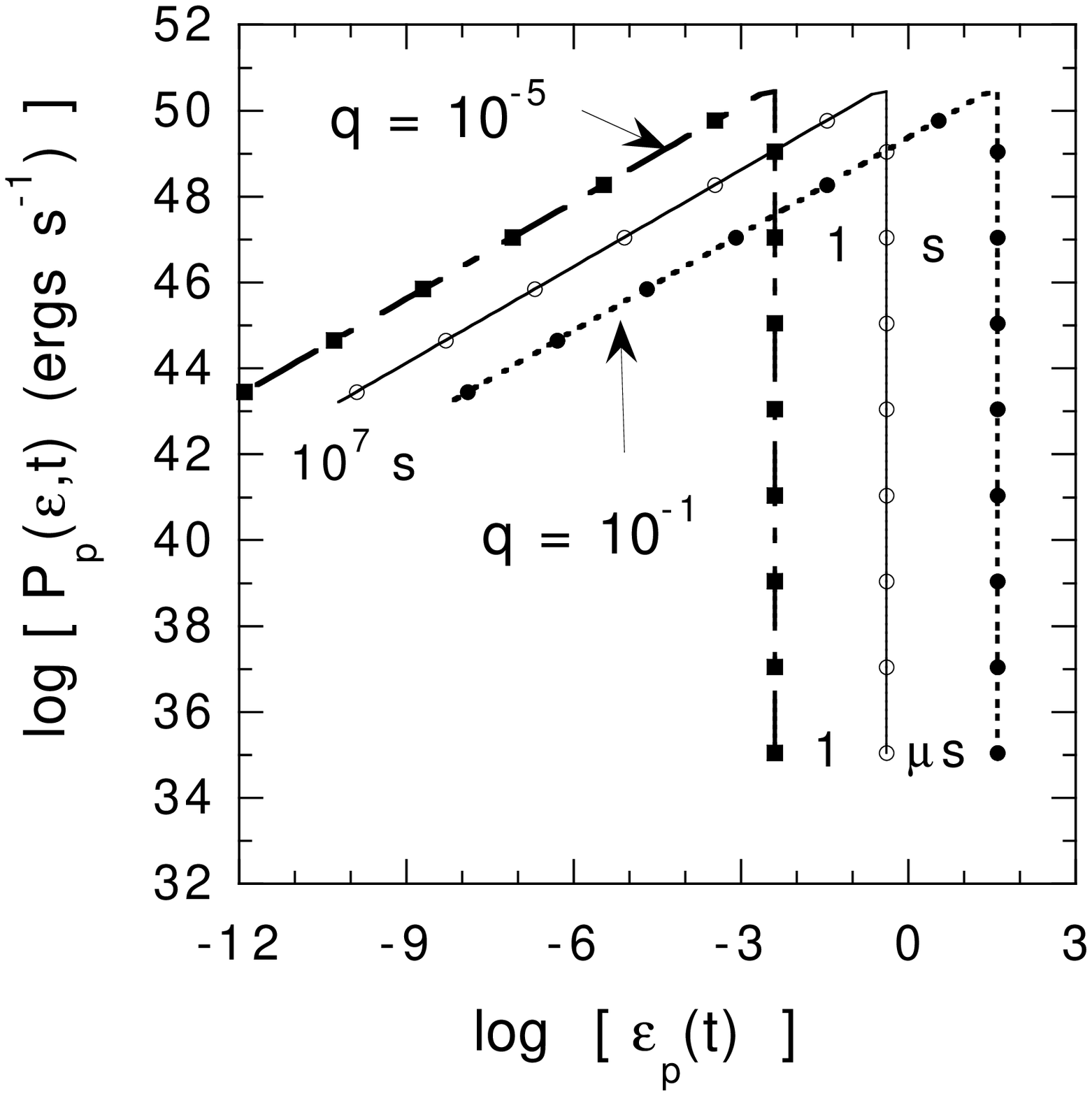}
\caption[]{(d) Same as Fig.\ 2a, but for different values of the equipartition
factor $q$. }
\end{figure}

\setcounter{figure}{1}
\begin{figure}
\epsfysize=12cm
\epsffile[0 20 500 500]{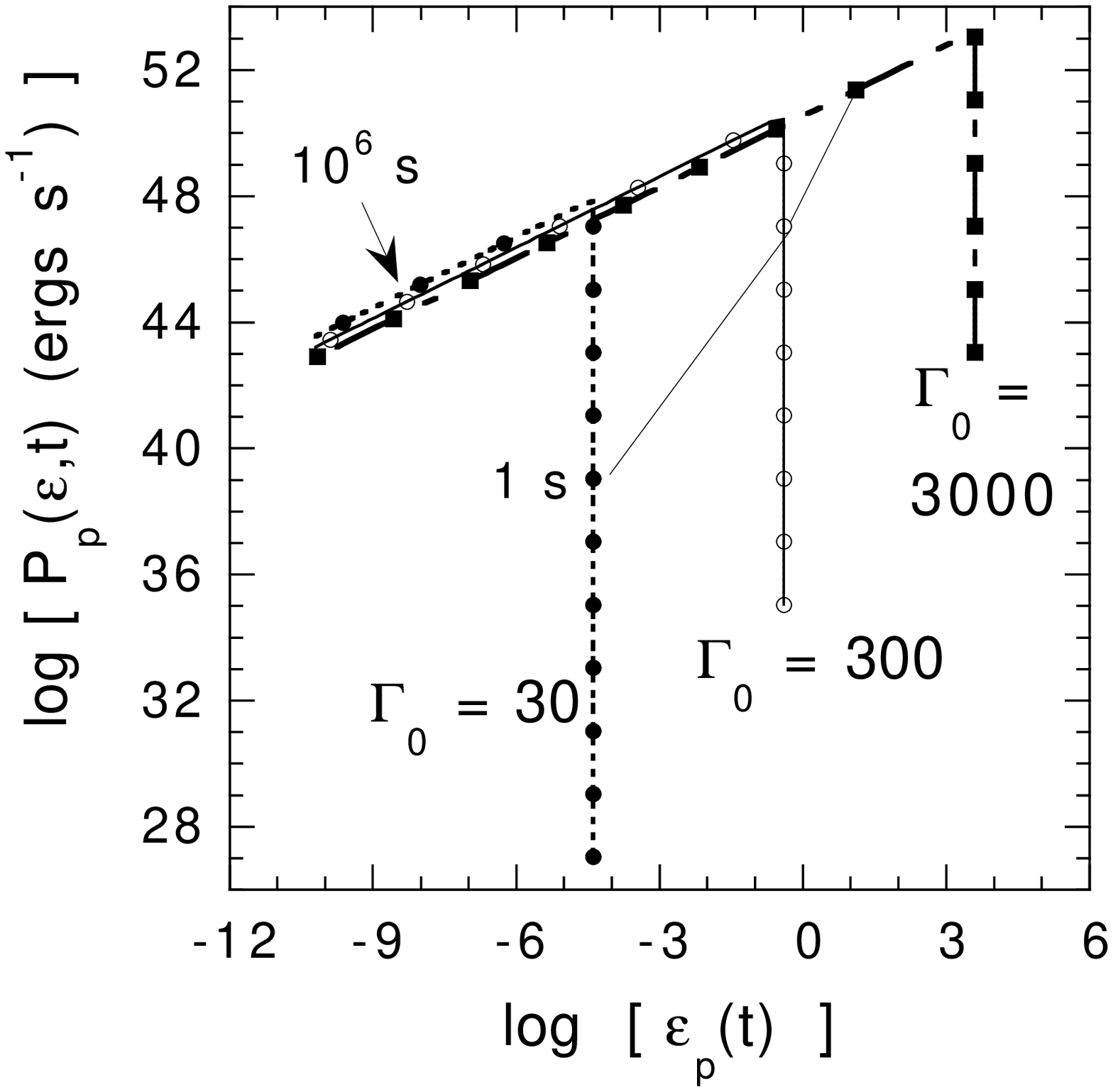}
\caption[17]{(e) Same as Fig.\ 2a, but for  different values of the baryon-loading factor
$\Gamma_0$. Note the strong dependence of the duration, peak power, and photon energy
of the $\nu L_\nu$ peak on $\Gamma_0$.}
\end{figure}

\setcounter{figure}{2}
\begin{figure}
\epsfysize=10cm
\rotate[r]{\epsffile[50 50 600 500]{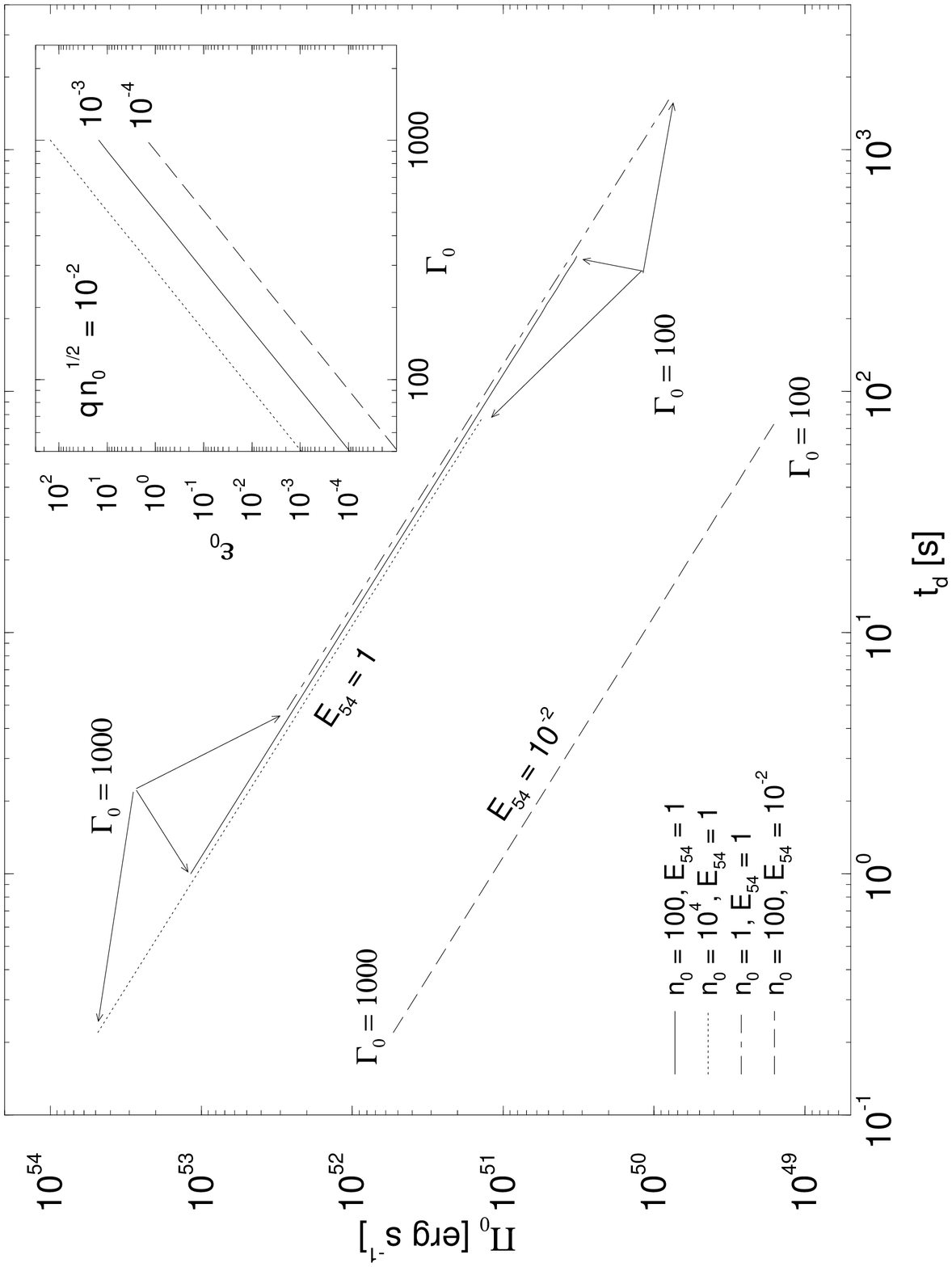}}
\caption[18]{Dependence of the observables $\Pi_0$, $t_d$
and ${\cal E}_0$ on the baryon-loading parameter $\Gamma_0$,
the total explosion energy $E_0 = 10^{54} E_{54}$ ergs and 
the external density $n_0$~(cm$^{-3}$), for $\Gamma_0$ 
ranging from 100 to 1000.
Other parameters are given in Table 2.}
\end{figure}

\end{document}